\documentclass[aps,prl,floatfix,twocolumn,nofootinbib]{revtex4}

\usepackage{amssymb}
\usepackage{amsmath}
\usepackage{amsfonts}
\usepackage{graphicx}
\usepackage{color}
\usepackage{xspace}
\usepackage{ulem}

\newcommand{\AddrAHEP}{
AHEP Group, Instituto de F\'{\i}sica Corpuscular -- 
C.S.I.C./Universitat de Val{\`e}ncia \\
Edificio de Institutos de Paterna, Apartado 22085,
E--46071 Val{\`e}ncia, Spain
}

\newcommand{\AddrUCL}{
Department of Physics and Astronomy, University College London,\\
London WC1E 6BT, United Kingdom
}

\begin{document}


\title{Falsifying High-Scale Leptogenesis at the LHC}

\author{Frank F. Deppisch} \email{f.deppisch@ucl.ac.uk}\affiliation{\AddrUCL}
\author{Julia Harz} \email{j.harz@ucl.ac.uk}\affiliation{\AddrUCL}
\author{Martin Hirsch} \email{mahirsch@ific.uv.es}\affiliation{\AddrAHEP}


\begin{abstract}
\noindent 
Measuring a non-zero value for the cross section of any lepton number violating (LNV) process
would put a strong lower limit on the washout factor for the effective
lepton number density in the early universe at times close to the
electroweak phase transition and thus would lead to important
constraints on any high-scale model for the generation of the observed
baryon asymmetry based on LNV. In particular, for leptogenesis models
with masses of the right-handed neutrinos heavier than the mass scale
observed at the LHC, the implied large washout factors would lead to a
violation of the out-of-equilibrium condition and exponentially
suppress the net lepton number produced in such leptogenesis models.
We thus demonstrate that the observation of LNV
processes at the LHC results in the falsification of high-scale
leptogenesis models. However, no conclusions about the viability of 
leptogenesis models can
be drawn from the non-observation of LNV processes.

\end{abstract}

\maketitle

\section{Introduction}
The observed baryon asymmetry of the universe~(BAU), 
measured in terms of the baryon-to-photon number density ratio
\cite{Ade:2013zuv},
\begin{align}
\label{eq:etaBobs}
  \eta_B^\text{obs} = \left(6.20 \pm 0.15\right) \times 10^{-10},
\end{align}
provides evidence for physics beyond the Standard Model (SM) 
\cite{reviews}. A popular scenario for explaining the BAU
is through the mechanism of leptogenesis (LG)~\cite{FY}. In the classic
LG scenario, heavy right-handed neutrinos decay out of
equilibrium and produce a lepton asymmetry. Necessary ingredients for
this process to occur are the presence of $(B-L)$ and $CP$
violation. The produced lepton asymmetry is then rapidly converted into
the observed BAU by $(B+L)$-violating sphaleron
interactions~\cite{KRS}.

\begin{figure}[t]
\centering
\vskip-6mm
\includegraphics[clip,width=0.49\linewidth]{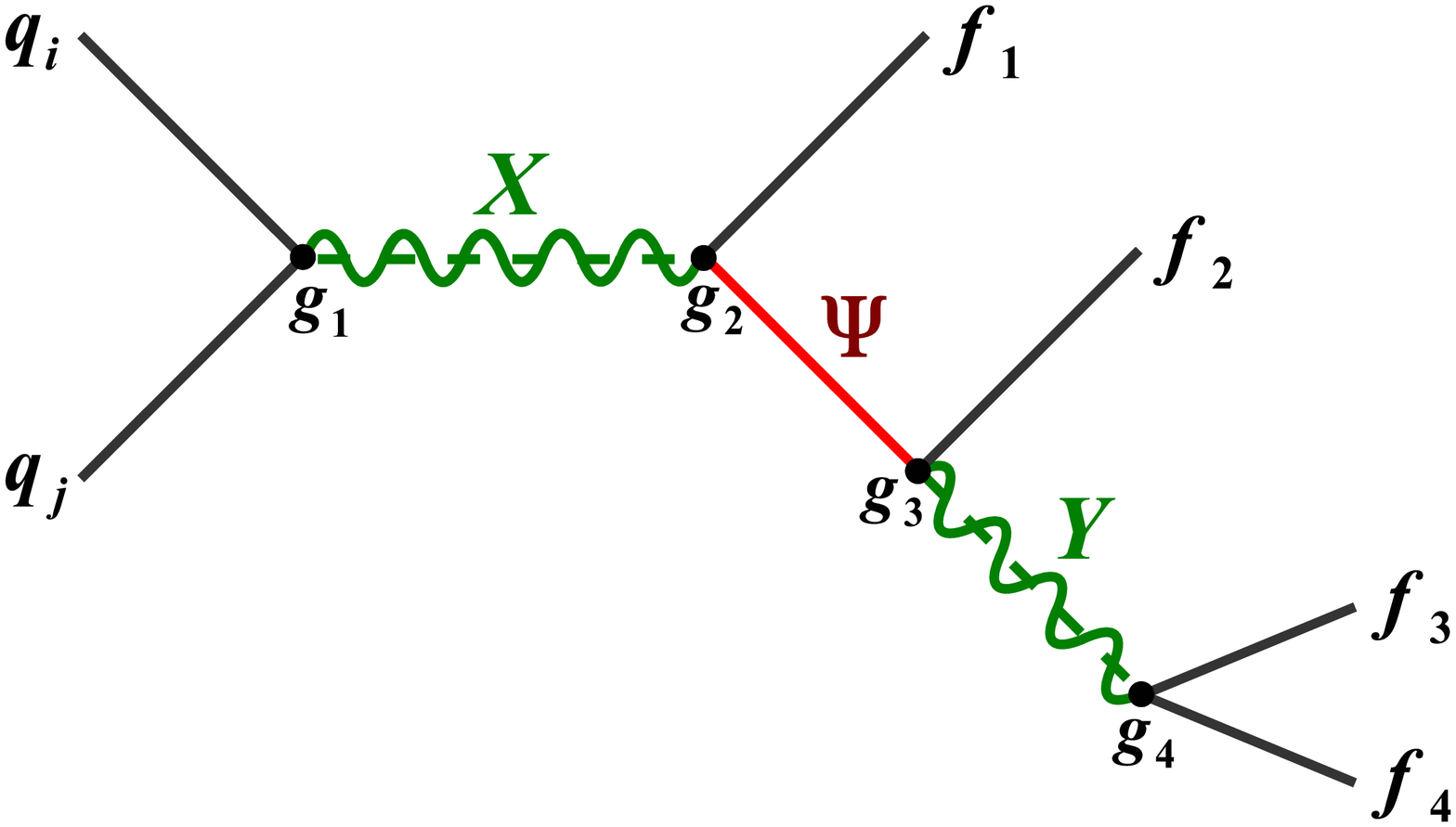}
\includegraphics[clip,width=0.49\linewidth]{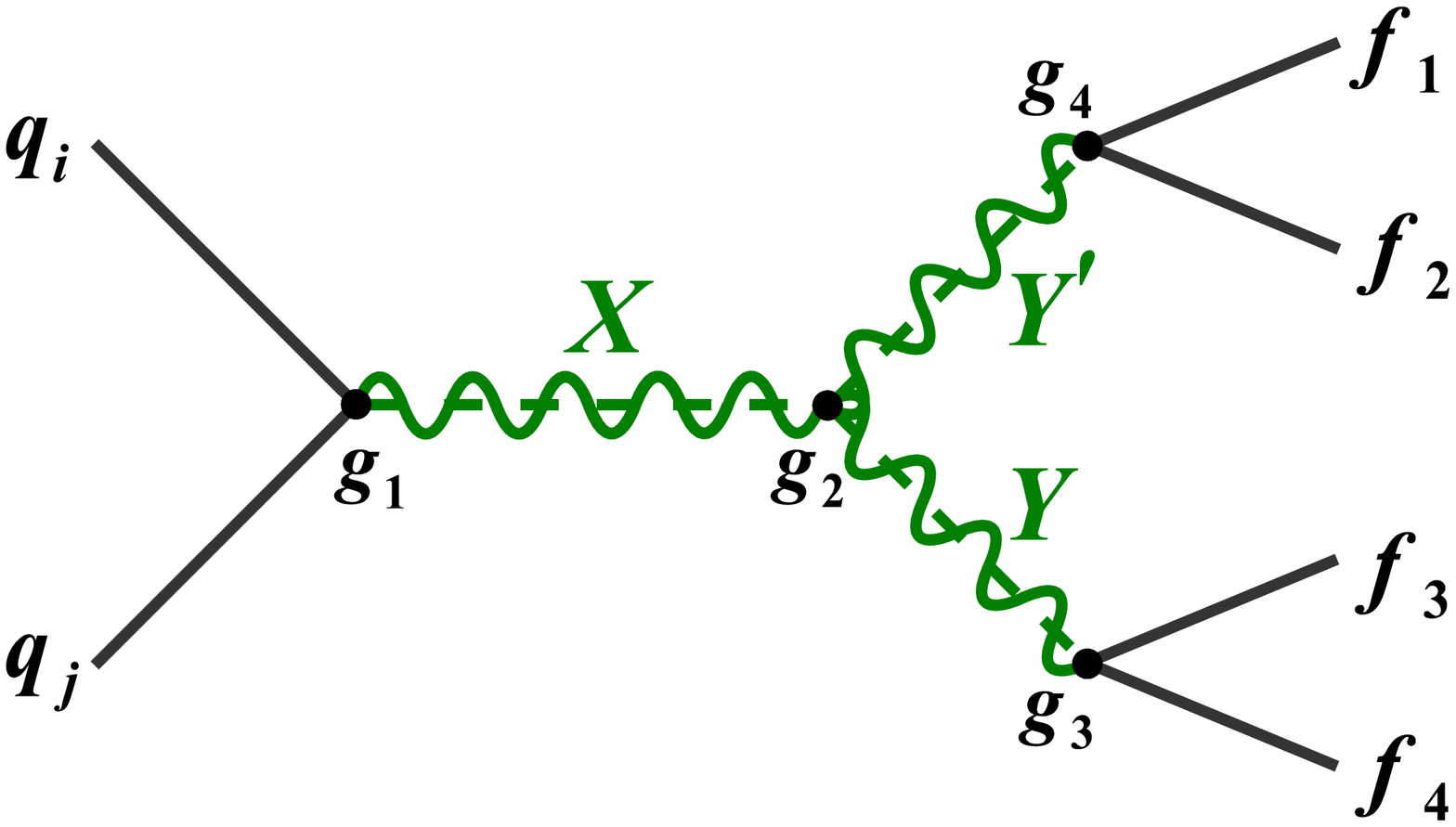}
\vskip-5mm
\caption{Possible diagrams contributing to the resonant same sign dilepton
signal $pp \to l^\pm
  l^\pm q q$ at the LHC. The intermediate particles $X$ and $Y^{(')}$ denote
different vector or scalar bosons, $\Psi$ indicates a fermion. In the general
case \cite{Bonnet:2012kh, Helo:2013ika}, any two of the four fermions $f_i$ can be leptons.}
\label{fig:decompositions}  
\end{figure}

Here, we consider lepton number violation (LNV) at the LHC through
same sign dilepton signals via resonant processes of the class shown
in Fig.~\ref{fig:decompositions}. The prototype example for this
signal is the resonant $W_R$ production in left-right symmetric
extensions of the SM with heavy Majorana neutrinos
\cite{Keung:1983uu}.  However, generic processes of this kind have
been discussed in \cite{Helo:2013ika} as tree level high-energy
completions of LNV operators that generate neutrinoless double beta
($0\nu\beta\beta$) decay \cite{Bonnet:2012kh}. For a recent review on
$0\nu\beta\beta$ decay see, for example \cite{Deppisch:2012nb}. In
Fig.~\ref{fig:decompositions}, the intermediate particles are
different vector or scalar bosons $X$ and $Y^{(')}$ and a fermion $\Psi$.
They decay to a final state with four SM fermions composed of two
quarks and two same sign charged leptons through unspecified
interactions of strengths $g_i$ ($i=1,2,3,4$). In the general case, any
combination of quark/anti-quark pairs $u$ and $d$ in the initial state
can be realized, and any two of the four final states $f_i$ can be leptons.
Note
that $0\nu\beta\beta$ constrains only final states with first
generation leptons, but at the LHC one could observe LNV in both
electrons and muons.

In this paper, we explore the consequences of the observation of LNV
 processes at the LHC on the viability of
LG mechanisms. Specifically, we discuss the impact of the
observation of LNV at the LHC on the rate of $\Delta L=2$ scattering
processes. As we will demonstrate, an observed non-zero cross section can be
converted into a lower limit on the washout of the lepton
asymmetry in the early universe. If the primordial lepton number asymmetry 
is originally generated above the LNV scale observed at the
LHC, the resulting washout will reduce the asymmetry exponentially, 
rendering LG ineffective.

We note that the question of falsifying LG at the LHC has
been investigated previously in reference \cite{Frere:2008ct} within
the context of the minimal left-right symmetric model. Our analysis
focuses instead on a model-independent approach in which we will derive
general limits from the hypothetical observation of the process 
$pp \to l^\pm l^\pm q q$.


\section{Same Sign Dileptons at the LHC}
Within the $\Delta L = 2$
resonant processes of the form $pp \to l^\pm l^\pm q q$, we focus on the s-channel
diagrams producing a scalar or vector boson $X$ resonantly which then
cascade decays to the final state $l^\pm l^\pm q q$ through on- or
off-shell decays. The parton level cross section can be approximated
by a Breit-Wigner resonance
\begin{align}
	\label{eq:cs_parton}
	\sigma(Q^2) = 
	\frac{4\pi}{9}(2J_{X}+1)
	\frac{\Gamma(X\to q_1 q_2)\Gamma(X\to 4f)}
	{(Q^2 - M_X^2)^2 + M_X^2 \Gamma_X^2},
\end{align}
with $J_X$ being the spin of the produced boson and $q_i$ indicating the
initial partons. 
The partial decay width $\Gamma(X\to 4f)$
describes the complete decay of $X$ as shown in Fig.~\ref{fig:decompositions}.
Integrating over the parton distribution
functions (PDFs) in narrow-width approximation of the resonance
\eqref{eq:cs_parton} yields the total LHC cross
section~\cite{Leike:1998wr}
\begin{align}
	\label{eq:cs_lhc}
	\sigma_\text{LHC} &=
	\frac{4\pi^2}{9 s}(2J_X+1)
	\frac{\Gamma_X}{M_X}
	f_{q_1 q_2}\left(\frac{M_X}{\sqrt{s}} , M_X^2 \right) \nonumber\\
  &\phantom{=}\,\times \text{Br}(X\to q_1 q_2)\text{Br}(X\to 4f),
\end{align}
with the LHC center of mass energy $\sqrt{s} = 14$~TeV and
\begin{align}
	\label{eq:pdf_integral}
	f_{q_1 q_2}\left(r, M^2 \right) = 
	\int_{r^2}^1 \frac{dx}{x}
	(
	   &q_1(x,M^2)q_2(r^2/x,M^2) + \nonumber\\
	   &q_2(x,M^2)q_1(r^2/x,M^2)
	).
\end{align}
Here, $q_i(x,Q^2)$ is the PDF of parton $q_i$ at momentum fraction $x$
and momentum transfer $Q^2$. For masses $M\approx 1-5$~TeV, this integral can be
well approximated as exponentially decreasing with
$M/\sqrt{s}$~\cite{Leike:1998wr},
\begin{align}
\label{eq:pdf_approximation}
	f_{q_1 q_2}\left(\frac{M}{\sqrt{s}}\right) \approx 
	A_{q_1 q_2}\times \exp\left(- C_{q_1 q_2} \frac{M}{\sqrt{s}}\right),
\end{align}
where the coefficients $A_{qq}$ and $C_{qq}$ depend on the combination of the
relevant
partons $q_1, q_2$, ranging between $A_{\bar u\bar u} \approx 200$ to
$A_{uu} \approx 4400$ and $C_{uu} \approx 26$ to $C_{\bar d\bar d}
\approx 51$.


\section{Leptogenesis} The relevant Boltzmann equations for leptogenesis
can be generically written in terms of the heavy neutrino and $(B-L)$
number densities per co-moving volume \cite{Giudice:2003jh} as 
function of its decay rate $\Gamma_D$, the CP asymmetry 
$\epsilon$ and the scattering rate $\Gamma_W$, which contains inverse $N$ 
decays as 
well as any other $\Delta L=1,2$ processes.

The scattering rate $\Gamma_W$ induced by the process $q q
\leftrightarrow l^\pm l^\pm q q$ is calculated from the reaction
density \cite{Giudice:2003jh}
\begin{align}
\label{eq:reaction_density}
  \gamma(q q \leftrightarrow l^\pm l^\pm q q) = 
  \frac{T}{32\pi^4}\int_0^\infty \!\! ds \,\, s^{3/2} \sigma(s) 
	K_1\left( \frac{\sqrt{s}}{T} \right),
\end{align}
with the $n$th-order modified Bessel function $K_n(x)$. Here, the process cross
section is not averaged over the initial particle quantum numbers.
Based on the same underlying process, the washout rate $\Gamma_W / H =
(\gamma/n_\gamma) / H$ and the LHC cross section $\sigma_\text{LHC}$ are
directly related. The
equilibrium photon density $n_\gamma \approx 2T^3/\pi^2$ and the
Hubble parameter $H \approx 1.66 \sqrt{g_*}T^2/M_\text{P}$ are
temperature dependent, with the effective number of relativistic
degrees of freedom $g_*$ ($\approx 107$ in the SM) and the Planck mass
$M_\text{P} = 1.2 \times 10^{19}$~GeV. This results in
\begin{align}
\label{eq:washout_factor_related}
  \frac{\Gamma_W}{H} &= 
  \frac{0.028}{\sqrt{g_*}}
  \frac{M_\text{P}M_X^3}{T^4}
  \frac{K_1\left( M_X/T \right)}
  {f_{q_1 q_2}\left( M_X / \sqrt{s} \right)}
  \times(s\sigma_\text{LHC}),
\end{align}
a relation independent of the branching ratios of the particle $X$ and
therefore valid for all coupling strengths $g_i$ and also independent
of the potential presence of other, lepton number conserving decay
modes. Evaluated at $T = M_X$, i.e. the approximate onset of the
washout process, Eq.~\eqref{eq:washout_factor_related} yields the
order of magnitude estimation
\begin{align}
\label{eq:washout_factor_estimation}
  \log_{10}\frac{\Gamma_W}{H} &\gtrsim
  6.9 + 0.6\left( \frac{M_X}{\text{TeV}} - 1 \right) +
\log_{10}\frac{\sigma_\text{LHC}}{\text{fb}},
\end{align}
using the conservative values $A_{qq} = 5000$ and $C_{qq}=26$ for
Eq.~\eqref{eq:pdf_approximation}. From this approximation alone it is clear that
the observation
of the resonant process $pp \to l^\pm l^\pm q q$ at the LHC
corresponds to a very strong washout of the lepton asymmetry in the
early universe. For example, the observation of a resonance at $M_X
\approx 2$~TeV with a cross section $\sigma_\text{LHC} \approx 1$~fb
corresponds to $\Gamma_W/H \approx 3\times 10^7$.
\begin{figure}[t]
\centering
\includegraphics[clip,width=0.90\linewidth]{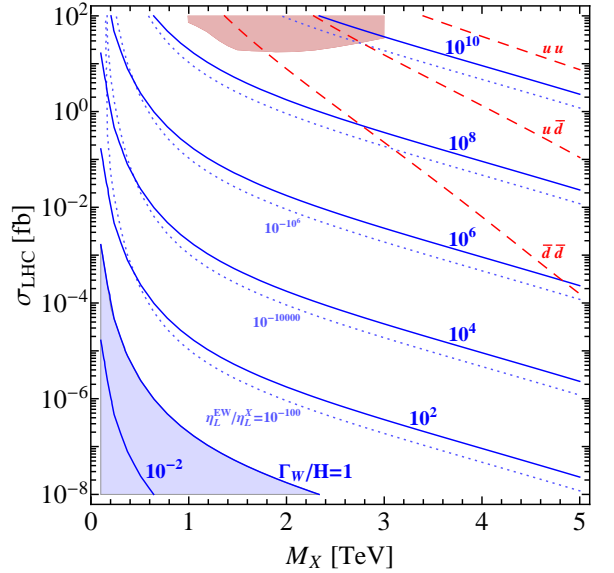}
\caption{Washout rate $\Gamma_W/H$ at $T = M_X$ as a function of $M_X$
  and $\sigma_\text{LHC}$ (solid blue contours). The dotted light blue 
	contours denote the surviving lepton asymmetry at the EW
	scale relative to its value at $M_X$, 
	$\eta_L^\text{EW}/\eta_L^X$.
	The red dashed curves are typical cross sections of the process 
	$pp \to l^\pm l^\pm q q$. The red shaded region at the top is excluded
due to recent searches for resonant same sign dileptons at the
LHC~\cite{CMS:2012uaa}.
}
\label{fig:GammaW_mX_sigma}
\end{figure}
The exact relation
\eqref{eq:washout_factor_related} is shown in
Fig.~\ref{fig:GammaW_mX_sigma}, based on the smallest washout
among all parton combinations. For any realistic cross section
observable at the LHC with $\sigma_\text{LHC}\gtrsim 10^{-2}$~fb, the
resulting lepton number washout in the early universe is always highly
effective ($\Gamma_W/H \gg 1$). The dashed curves, for example, show typical
cross sections for different
parton combinations in the case of a particle $X$ with gauge-strength
total width, $\Gamma_X/M_X = g^2/(32\pi)$, with $g=0.5$ and branching ratios
$\text{Br}(X\to q_1 q_2) = \text{Br}(X\to 4f) = 0.5$. 
Without a source regenerating the lepton asymmetry below $M_X$,
the washout leads to an exponential suppression of the surviving
asymmetry at the EW scale compared to the value present at $M_X$, 
$\eta_L^\text{EW}/\eta_L^\text{X} \approx \exp(-\Gamma_W/H)$. This suppression 
is also shown in Fig.~\ref{fig:GammaW_mX_sigma}, highlighting that the 
observation of LNV processes at the level $\sigma_\text{LHC} \gtrsim 10^{-2} 
\text{ fb}$ would necessarily result in an enormous washout of any 
pre-existing lepton asymmetry.

It should be stressed again that the above analysis is highly model
independent and purely based on the observables $M_X$ and
$\sigma_\text{LHC}$ of the process. The approximations used in our 
calculation, such as the
narrow-width resonance assumption, are not expected to change this
conclusion in any way. In fact, the direct relation between the LHC cross
section and the washout rate is expected to hold for any LNV process, with
a proportionality only affected by the kinematics of the process.

In order to further assess the impact of the resulting washout, we calculate
the baryon asymmetry in the standard LG scenario with one
heavy neutrino $N$, neglecting all other washout reactions. The
Boltzmann equations in this case are most
compactly expressed in terms of the out-of-equilibrium heavy Majorana
neutrino density deviation $\delta\eta_N =
\eta_N/\eta_N^{\rm eq} - 1$ and the lepton density $\eta_L =
n_L/n_\gamma$ normalized to the photon 
density~$n_\gamma$~\cite{Deppisch:2010fr},%
\begin{align}%
\label{eq:Boltzmann_compact_N}
	\frac{d\delta\eta_N}{dz} &=
	\frac{K_1(r_N z)}{K_2(r_N z)}
	\bigg[ r_N + \Big(1 - r_N^2 K_D z \Big)\delta\eta_N\bigg], \\
\label{eq:Boltzmann_compact_L}
	\frac{d\eta_L}{dz} &=
	 \epsilon K_D r_N^4 z^3 K_1(r_N z)\delta\eta_N  - K_W z^3 K_1(z)\eta_L,
\end{align}%
with the decay factor $K_D = \Gamma_D / H$ at $T=M_N$ and the washout
factor $K_W = \Gamma_W / H$ at $T=M_X$. We define the evolution
parameter $z=M_X/T$. The above Boltzmann equations explicitly contain the
full temperature ($z$) dependence with the hierarchy
between the masses of $N$ and $X$ given by $r_N = M_N/M_X$.  Note
that Eqs.~\eqref{eq:Boltzmann_compact_N} and
\eqref{eq:Boltzmann_compact_L} implicitly assume that the $CP$
asymmetry is generated from the decay of the heavy neutrino. Other
mechanisms will yield similar results.

\begin{figure}[t]
\centering
\includegraphics[clip,width=0.90\linewidth]{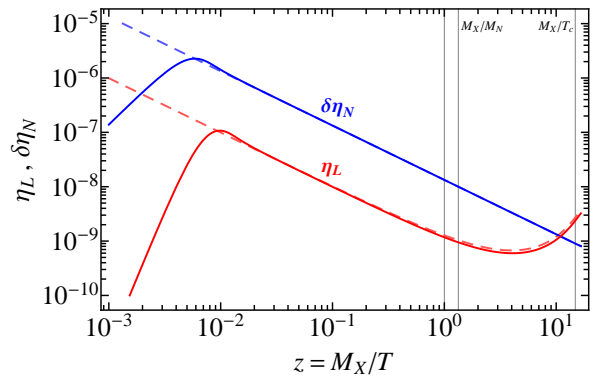}
\caption{Solution to the Boltzmann equations for $M_N=1.5$~TeV,
$M_X=2$~TeV, $CP$ asymmetry $\epsilon =
  10^{-2}$, $K_D = 10^8$ and $K_W \approx 6 \times 10^6$ (smallest
  washout corresponding to $\sigma_\text{LHC} = 0.1$~fb). The dashed
  lines denote approximated solutions discussed in the text.}
\label{fig:boltzmann}
\end{figure}
Fig.~\ref{fig:boltzmann} shows a typical solution to the Boltzmann
equations \eqref{eq:Boltzmann_compact_N} and
\eqref{eq:Boltzmann_compact_L} with $M_N=1.5$~TeV, $M_X=2$~TeV, $CP$
asymmetry $\epsilon = 10^{-2}$, $K_D = 10^8$, and $K_W \approx 6 \times
10^6$ (smallest washout corresponding to $\sigma_\text{LHC} =
0.1$~fb). For sufficiently strong rates $K_D$ and $K_W$, the linear
drop-off behaviour depicted can be approximated as $\delta\eta_N = 1 /
(r_N K_D z)$ and $\eta_L = r_N^2 \epsilon / (K_W z)$, respectively, in
which case $\eta_L$ is generally independent of $K_D$
\cite{Deppisch:2010fr}. For $M_N = M_X$, $\eta_L$ will eventually
freeze-out to a constant value, but for a different hierarchy, the
contributions to $\eta_L$ depend differently on the temperature. In
the case $r_N<1$, shown in the figure, heavy neutrino decays can not be 
compensated by the washout for $z > 1$,
whereas for $r_N>1$ the washout is always effective for $z>1$ leading
to an over-proportional drop-off. The general behaviour can be well
approximated by $\eta_L \approx r_N^2 \epsilon / (K_W
z)\exp((1-r_N)z)$. In Fig.~\ref{fig:boltzmann}, this approximation
is shown as a red dashed curve. Such a behaviour is not realistic in a
given model where other processes are expected to contribute, but
allows to draw model-independent conclusions. Most importantly we
neglect washout from processes mediated by the heavy neutrino driven
by the same interaction(s) that generate the lepton asymmetry. Taking
into account all washout processes in a consistent fashion will
guarantee a well-behaved freeze-out behaviour for $\eta_L$. In this
spirit, our solution for $\eta_L$ can be considered as a conservative,
model-independent but also possibly weak upper limit on the generated
lepton asymmetry.

The conversion of the lepton number to the final ba\-ryon asymmetry can
be calculated by $\eta_B = - d_\text{rec} r_{B/L} \eta_L(T_c)$ with 
$r_{B/L} = (8N_g + 4N_H)/(14N_g + 9N_H) \approx 1/2$
in a general theory with $N_g$ fermion generations and $N_H$ Higgs doublets
\cite{Khlebnikov:1988sr}. The critical temperature of the
electroweak phase transition is denoted by $T_c \approx 135$~GeV, 
$d_\text{rec} \approx 1/27$ (in the SM) describes the increase of the photon
density during the recombination epoch, and $\eta_L(T_c)$ is the lepton
asymmetry at the sphaleron decoupling temperature. For details see
\cite{Pilaftsis:2005rv}.

\begin{figure}[t]
\centering
\includegraphics[clip,width=0.85\linewidth]{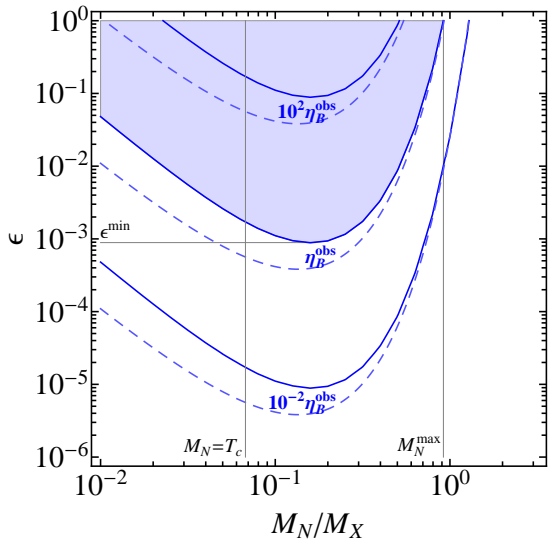}
\caption{Baryon asymmetry $\eta_B$ as a function of $M_N/M_X$ and
  $\epsilon$ for $M_X = 2$~TeV and $\sigma_\text{LHC} = 0.1$~fb (solid
  contours). The intermediate contour corresponds to the observed
  value $\eta_B^\text{obs}$, the other two contours give
  100 times higher and lower values, respectively. Correspondingly,
  the dashed contours are determined using the approximation
  Eq.~\eqref{eq:etaBapprox}.}
\label{fig:etaB_rN_epsilon}
\end{figure}
We arrive at an upper limit for the baryon asymmetry,
\begin{equation}
  \label{eq:etaBapprox}
	|\eta_B| \lesssim 	
	\frac{M_N^2}{M_X^2} \frac{T_c}{M_X} 
	\frac{r_{B/L} d_\text{rec} |\epsilon|}
	{K_W(M_X,\sigma_\text{LHC})}
	e^{(M_X-M_N)/T_c}
\end{equation}
that can be compared to the observed value \eqref{eq:etaBobs}. By using the
derived approximation \eqref{eq:washout_factor_estimation}, we can set an
upper limit 
on the baryon asymmetry as a function of the LG parameters $m_N$ 
and $\epsilon$, and the observables $M_X$ and $\sigma_\text{LHC}$ of the 
LHC process,
\begin{align}
  \label{eq:etaBestimation}
	\log_{10} \left|\frac{\eta_B}{\eta_B^\text{obs}}\right| &\lesssim 
	2.4\, \frac{M_X}{\text{TeV}} \left( 1 - \frac{4}{3} \frac{M_N}{M_X}
\right)\nonumber \\
&+ \log_{10} \left[ |\epsilon|\, \left(
\frac{\sigma_\text{LHC}}{\text{fb}} \vphantom{\frac{4}{3}
\frac{M_N}{M_X}}\right)^{-1} \left( \frac{4}{3} \frac{M_N}{M_X} \right)^2
\right].
\end{align}
In Fig.~\ref{fig:etaB_rN_epsilon} the resulting baryon asymmetry for both the
exact solution of the Boltzmann equations (solid line) as well as the
approximation \eqref{eq:etaBobs} (dashed line) is shown as a
function of $r_N = M_N/M_X$ and $\epsilon$ for $M_X = 2$~TeV and
$\sigma_\text{LHC} = 0.1$~fb. Two
important conclusions can be drawn: (i) For $M_N >
M_N^\text{max} \approx M_X$ it is not possible to generate a large
enough baryon asymmetry. As our calculation gives a conservative upper
limit for $\eta_B$, this means that the observation of the LNV process
at the LHC excludes high energy LG models. (ii) For $M_N <
M_X$ there exists a lower limit on the $CP$ asymmetry $\epsilon >
\epsilon^\text{min} \approx 10^{-3}$, which strongly
constrains resonant LG models.

\section{Discussion and conclusions} We have discussed the impact of a
possible observation of lepton number violation at the LHC on
leptogenesis. We have shown that for right-handed neutrinos heavier
than the mass scale at which LNV is observed at the LHC, the
resulting washout factor will reduce any pre-existing lepton
asymmetry $L$ exponentially, rendering LG ineffective. Our arguments
should be generally valid and not depend on the particular realization
of LG, although we have concentrated on the ``standard'' scenario with
right-handed neutrinos.
Thus, high-scale thermal leptogenesis models can be
falsified in case of the observation of LNV processes at the LHC. However, it
should be stressed that no conclusions about the viability of LG can be drawn
from the non-observation of these processes.

There are a few possible caveats to consider: 
(i) One could imagine a situation where LNV is generated in the early universe
in the third family only. As there is no experimental proof so far
that $e^{\pm}e^{\pm} \leftrightarrow \tau^{\pm}\tau^{\pm}$ was in
equilibrium in the early universe, the observation of LNV at the LHC for only
$e$ and/or $\mu$ is not sufficient in this case. A non-zero
observation of $pp \to l^\pm l^\pm q q$ for either $ll = e e$, $\mu\mu$ and
$\tau\tau$, or for $e\mu$ and $e(\mu)\tau$ is necessary to unambiguously 
falsify LG models in the way presented.
(ii) Our discussion leaves open the possibility of LG with $M_N < M_X$. 
Resonant sub-TeV scale LG with large $CP$ asymmetry $\epsilon$, for example, 
would still be possible. This loophole is not limited to the
classical LG with right-handed neutrinos.  In general, any mechanism 
producing a ``hidden'' lepton number which is converted to $B$
below $M_X$, would not be ruled out. Nevertheless, it is still possible 
to derive a large model-independent lower limit on the required $CP$
asymmetry. 
(iii) SM sphaleron processes only affect electroweak fermion doublets, 
but left- and right-handed fermions are
in thermal equilibrium around the electroweak scale for Yukawa couplings
larger than $\approx 10^{-8}$. This is the case for all charged leptons. 
Our conclusions therefore also apply if the LNV
process at the LHC involves right-handed leptons.

\begin{figure}[t]
\centering
\includegraphics[clip,width=0.45\linewidth]{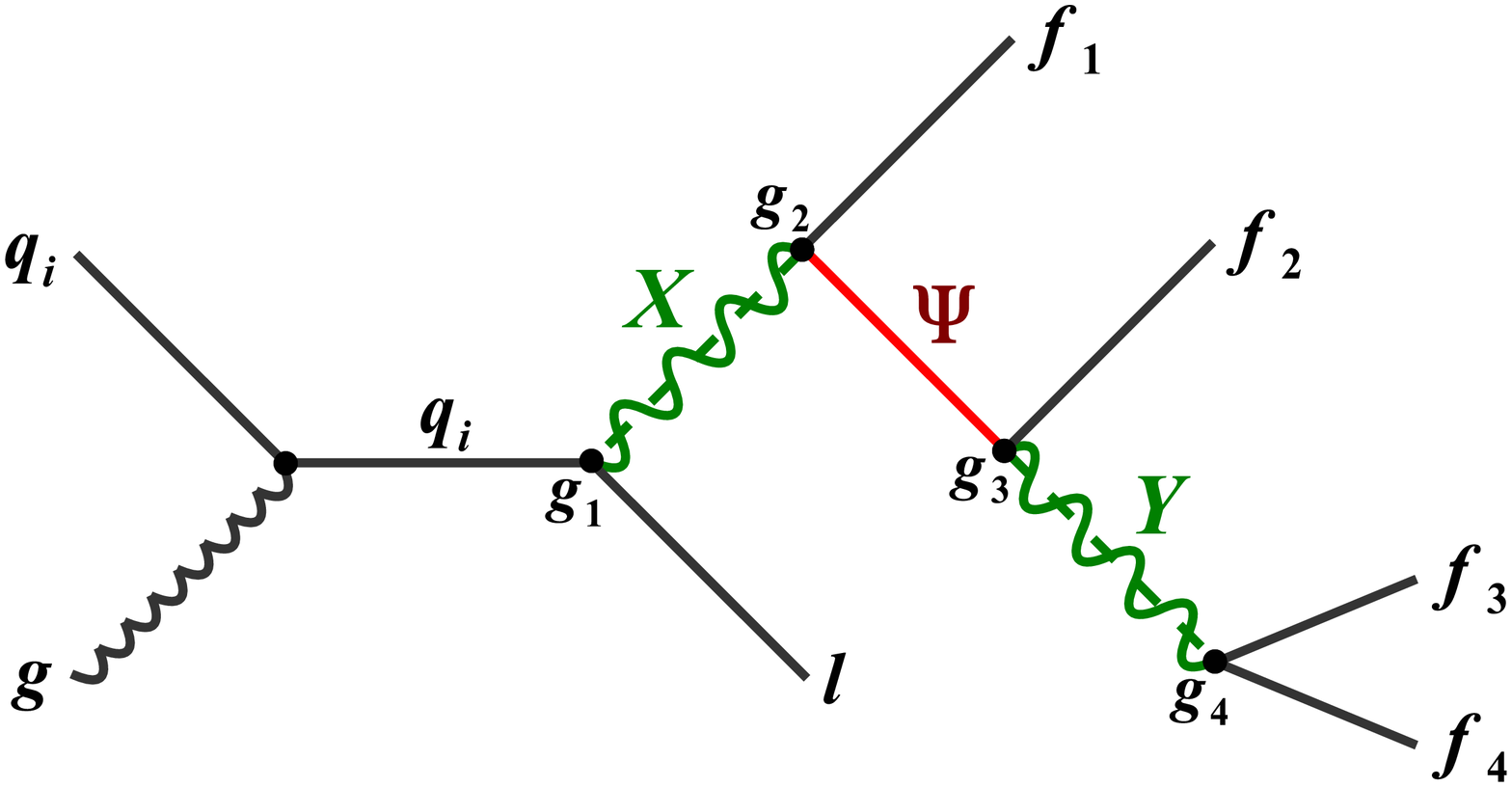}
\includegraphics[clip,width=0.45\linewidth]{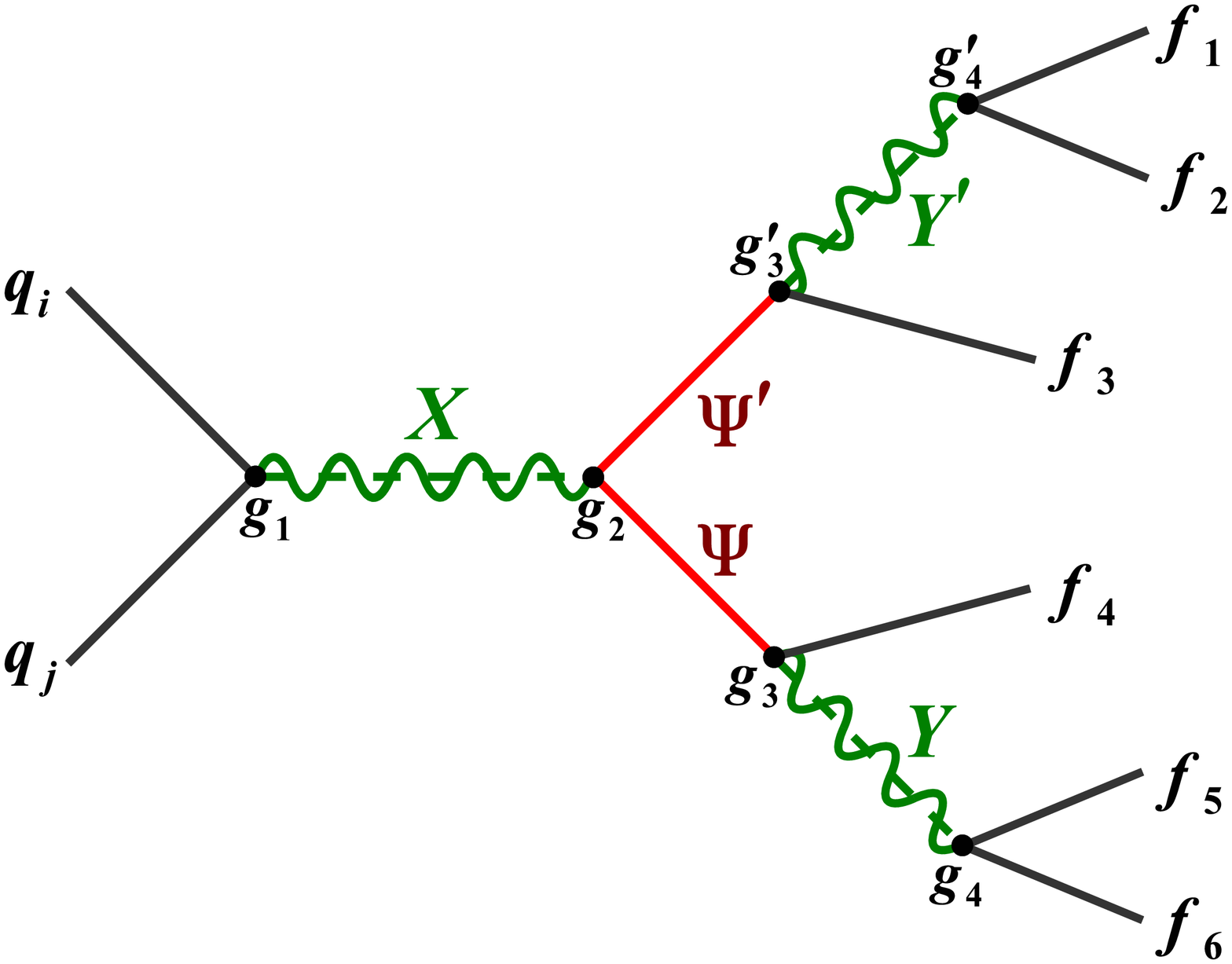}
\caption{Further example diagrams for different LNV process classes 
at the LHC with 5 (left) and 6 (right) final state particles. Note that 
in the latter case, in addition to the $qq$ initial state shown, also 
gluon-gluon initial state diagrams can exist.}
\label{fig:decompositions2}  
\end{figure}

Finally, although we concentrated on the resonant process of
Fig.~\ref{fig:decompositions}, we believe our argumentation can be easily
adapted to other cases. For example, leptoquark production can occur
as in Fig.~\ref{fig:decompositions2}~(left) and pair production of
heavy states as in Fig.~\ref{fig:decompositions2}~(right). We conjecture 
that our relation between the observation of the LNV at LHC and the 
lower limit on the washout factor is valid for these cases too, 
most likely even with larger numerical factors. The discovery of LNV at 
the LHC could then have profound implications, especially if combined
with the observation of $0\nu\beta\beta$ decay at a rate expected 
from a short range operator induced by a process as in 
Fig.~\ref{fig:decompositions}. Such an experimental scenario would 
dis-favour the standard high scale seesaw mechanism as both leptogenesis and
the dominance of light Majorana neutrinos mediating $0\nu\beta\beta$ decay 
are rendered ineffective.

\section{Acknowledgments}
\begin{acknowledgments}
The work of FFD and JH was supported partly by the London Centre for
Terauniverse Studies (LCTS), using funding from the European Research
Council via the Advanced Investigator Grant 267352. The work of MH was
supported by the Spanish MINECO under grants FPA2011-22975 and
MULTIDARK CSD2009-00064 (Consolider-Ingenio 2010 Programme), by
Prometeo/2009/091 (Generalitat Valenciana), and by the EU ITN UNILHC
PITN-GA-2009-237920. FFD and JH would like to thank Robert Thorne for
useful discussions.
\end{acknowledgments}

\end{document}